\documentclass[journal]{IEEEtran}
\usepackage{amsmath}
\usepackage{graphicx}
\usepackage{siunitx}

\usepackage{cite}

\ifCLASSINFOpdf
\else
\fi
\usepackage{xcolor}
\begin{document}
%
\title{Physical Design and Experimental Verification of a Huygens' Metasurface Two-lens System for Phased-array Scan-angle Enhancement}
%
%
%

\author{Jaemin~Kim,~\IEEEmembership{Student Member,~IEEE,}~Gleb~A.~Egorov,
        and~George~V.~Eleftheriades,~\IEEEmembership{Fellow,~IEEE}
\thanks{The authors are with the Edward S. Rogers Sr. Department of Electrical and Computer Engineering, University of Toronto, Toronto, Canada (e-mail: jmn.kim@mail.utoronto.ca; gelefth@waves.utoronto.ca).}}

%
%

\markboth{}%
{Shell \MakeLowercase{\textit{et al.}}: Bare Demo of IEEEtran.cls for IEEE Journals}
%



\maketitle

\begin{abstract}
Over the past decades, many radome designs to extend the angular scan range of phased-array antennas have been devised by utilizing dielectric materials and metamaterials.
More recently, metasurface technology such as planar lenses and beam deflectors have been applied to phased arrays, enabling scan-angle enhancers to have a low profile.
In this work, a physical Huygens' metasurface (HMS) two-lens system for scan-angle doubling of a phased array is presented.
For the HMS unit cells, the wire-loop topology is deployed to achieve high transmission for the required phase-angle shift.
The proposed two-lens system is analyzed by full-wave simulations and experiments.
The simulation results demonstrate that the scan angle doubles when the incident angle is below $\ang{15}$ in accordance to the design specification.
Furthermore, the directivity degradation of the refracted beams by the two-HMS lenses is in good agreement with theory.
Finally, a fabricated two-lens system with two $15\lambda$ long by $15\lambda$ wide metasurface lenses and a $16\times16$-element patch antenna array as a source is experimentally verified at $\SI{10}{\GHz}$.
The experimental results are in good agreement with the simulated results by showing angle-doubling performance with $\ang{\pm2}$ scan errors.
\end{abstract}


%
\IEEEpeerreviewmaketitle

\section{Introduction}
%
%
%
%
\IEEEPARstart{P}{hased} array antennas are being used in mobile and satellite applications to scan a beam electronically~\cite{Mailloux:handbook}.
However, the scanning range of a phased array can be limited due to various issues such as directivity degradation and the presence of grating lobes for larger element spacings such as the case of sub-arraying~\cite{Mailloux:handbook}.
Much research has been conducted in the last decades to enhance the scan range of phased arrays.
A cylindrical dome antenna with a phase-incurring layer was one of the early theoretical solutions for  scan-angle enhancement~\cite{Steyskal:radome}.
A dielectric dome with a linear source array was realized experimentally to extend the range of scan angle by optimizing the shape and thickness of the dome using  geometrical optics~\cite{Kawahara:radome}.
Moreover, radomes based on recent research topics such as transformation optics and negative-index metamaterials have been studied~\cite{Sun:TO, Lam:NIM}.
Nevertheless, the thick profile of these radomes impede the realization of compact angle-enhancement systems. 

To reduce the thickness of the antenna architecture, metasurface technology has been applied to scan-angle enhancement of phased arrays~\cite{Benini:MTSlens,Gleb:single_lens,Gleb:two_lens}.
In~\cite{Benini:MTSlens}, a curve shaped metasurface dome (meta-radome) has been used to deflect the angle of the incident beam and improve the scan range of a phased array.
This approach reduces the volume of the angle enhancement system, but the phased-array elements need to be excited with non-uniform magnitudes and non-linear phase shifts for directive radiation of the beam through the meta-radome.

More recently, Huygens' metasurface (HMS) theory has been utilized for realizing  scan-angle enhancement systems~\cite{Gleb:single_lens,Gleb:two_lens}.
The HMSs are subwavelength thin sheets comprising scatterers or unit cells which contain co-located orthogonal electric and magnetic dipole moments~\cite{Vasileios:2021MTTS}.
HMSs have the ability to refract a given incident field to a desired transmitted field -- the particular incident and transmitted fields are chosen during the design of a particular HMS. The discontinuity in the fields is supported by the electric and magnetic surface current densities which are excited by the incident electromagnetic field in steady state~\cite{Vasileios:2021MTTS}. 
Thus, they can provide the desired spatially-varying phase shift between the incident and transmitted fields.
The HMSs have been used in a wide range of applications such as reflectionless wide-angle refraction and beamforming with bianisotropic surfaces~\cite{Wong:2016AWPL, Chen:2018PRB, Vasileios:2022IEEE_Magazine}. 
More recently, HMSs have demonstrated the ability of independenet 
phase and amplitude control for precise antenna beamforming through the excitation of auxialiary surface waves~\cite{Epstein:2016PRL, Vasileios:2021AWPL}.
In terms of extending the scan range of phased arrays, HMS planar lenses with ray optical theory were discussed in~\cite{Gleb:single_lens,Gleb:two_lens}.
Compared with the meta-domes and radomes in~\cite{Steyskal:radome, Kawahara:radome, Benini:MTSlens}, the structures with HMS lenses are planar.
Furthermore, the array elements can be linearly phased as usually.
However, these HMS lenses have been based on ideal impedance sheets.  

In this paper, we demonstrate a physical design, including losses, for a two-HMS lens system for angle doubling by full-wave simulations and experiments.
Here, the desired scan range of the angle enhancement system is from $\ang{-30}$ to $\ang{30}$, whereas the source array steers its beam electronically between $\ang{-15}$ and $\ang{15}$.
The HMS unit cells are designed with a wire-loop topology to exhibit high transmittance over all required phase angles.
Note that the stacked-layer unit cell topology is also a popular approach for designing HMSs by utilizing the equivalent transmission-line model~\cite{Vasileios:2021MTTS}.
However, the stacked-layer HMS unit cells can suffer from significant loss when the phase angle of $\mathrm{S}_{21}$ is near $\ang{0}$ due to resonance~\cite{Chen:Lens}. Hence the power transmission efficiency of HMSs can be compromised.
On the other hand, the wire-loop unit cells we utilize in this work exhibit high transmittance for the desired phase angles of $\mathrm{S}_{21}$ including $\ang{0}$ as shown with full-wave simulations.
Additionally, the scan angle of the lossy two-HMS scan-angle doubler is enhanced almost twice and with low scan error, when the incident beam angle is between $\ang{-15}$ and $\ang{15}$, according to the desired specification.
\begin{figure}[!t]
\centering
\includegraphics{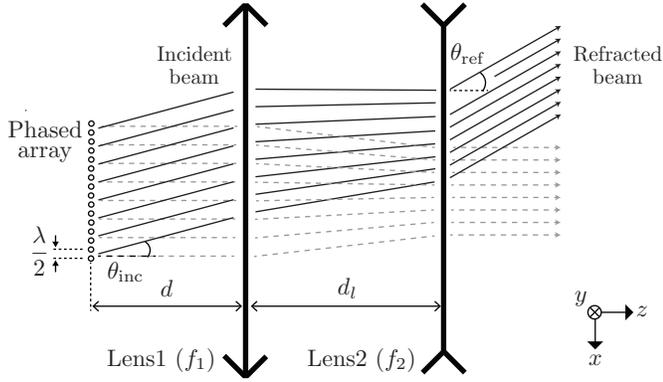}
\caption{A schematic of the proposed two-HMS lens angle-enhancing system with coordinate system vectors and angles for the incident and refracted beams. Parameters such as distances and focal lengths are specified. The dotted and solid ray paths of the system depict beam scanning when the incident beam angle is $\ang{0}$ and $\theta_{\mathrm{inc}}$, respectively.}
\label{fig1}
\end{figure}

\section{Two-lens system for scan enhancement}
Two-HMS lens scan-angle enhancement is achieved with a converging lens and a diverging lens which are placed in the near-field region of a source array as depicted in Fig.~\ref{fig1}~\cite{Gleb:two_lens}.
Ray tracing through the two-lens system can be expressed by ray transfer matrix analysis~\cite{Yariv:photonics}.
The ray transfer matrix for a two-lens system is shown in~\eqref{eq1}~and~\eqref{eq2}, which gives the position and angle of a ray when passing through the lenses.
\begin{equation} \label{eq1}
    \begin{bmatrix}
x^\prime\\ \theta_{\mathrm{ref}}
\end{bmatrix}
=
\begin{bmatrix}
A&B\\C&D
\end{bmatrix}
\begin{bmatrix}
x\\ \theta_{\mathrm{inc}}
\end{bmatrix},
\end{equation}
where A, B, C, and D are given by

\begin{equation} \label{eq2}
\begin{split}
    &A=1-\frac{d_l}{f_1}\\
    &B=d\left(1-\frac{d_l}{f_1}\right) + d_l\\
    &C=\frac{1}{f_1 f_2}(d_l-f_1-f_2)\\
    &D = 1-\frac{d_l}{f_2} + \frac{d}{f_1f_2}(d_l-f_1-f_2),\\
\end{split}
\end{equation}
where $d$ is the distance between a source and the first lens, $d_l$ is the distance between the two lenses, and $f_1$ and $f_2$ are the focal lengths of each lens.
For the two-lens system to work as a scan enhancer, $C$ and $D$ in the transfer matrix must satisfy the condition in~\eqref{eq3}.
\begin{equation} \label{eq3}
\begin{split}
    &d_l-f_1-f_2=0\\
\end{split}
\end{equation}
As a result, the desired angle of a ray passing through the two-lens system can be obtained by~\eqref{eq4}, where $\alpha$ is the angular scan enhancement factor for the two-lens system.
\begin{equation} \label{eq4}
\begin{split}
    &\theta_{\mathrm{ref}} = \left(1-\frac{d_l}{f_2}\right)\theta_{\mathrm{inc}} = \alpha \theta_{\mathrm{inc}}\\
\end{split}
\end{equation}

To fulfill the requirement for the scan-angle doubler in the two-lens system, $\alpha$ has to be 2, and it leads to $f_1 = 2d_l=f_c$ and $f_2 =-d_l=f_d$, where the $f_c$ and $f_d$ are the focal lengths of the converging and diverging lens, respectively.
Here, the angle-doubling system takes $d_l=4\lambda$ at $\SI{10}{\GHz}$ resulting in $f_c = 8\lambda$ and $f_d = -4\lambda$ as shown in Fig.~\ref{fig1}.
The source array is located at a distance $d=4\lambda$ away from the first converging lens.
Note that the source array can be placed at any $d$ from the lens as the theory shows, but $d=4\lambda$ was chosen for our experimental convenience.
The array comprises  16-element infinitely long electric current line sources,  $\lambda/2$-spaced, to propagate transverse electric (TE) polarized fields. The array is phased to create off-broadside beams between $\ang{-15}$ and $\ang{15}$.

\section{Huygens' metasurface lens design}
To realize the scan-angle enhancer with the two lenses, we utilize Huygens' metasurface (HMS) lenses because the HMS unit cells can be designed with high magnitude of $\mathrm{S}_{21}$ over all required phase angles~\cite{Chen:Lens,Epstein:TAP2016}.    
The phase angles of $\mathrm{S}_{21}$ of the HMS lenses should be specified by~\eqref{eq5} as the quadratic phase profile for a lens~\cite{Gleb:single_lens},

\begin{figure}[!t]
\centering
\includegraphics{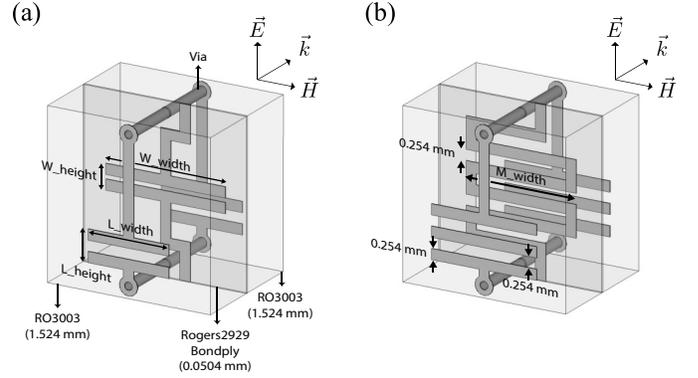}
\caption{Two kinds of wire-loop Huygens' metasurface unit cells and parameters used for S-parameter optimization. A loop structure is realized with printed capacitors on the outer layers in (a) and (b). For the wire structure in the middle layer, a printed capacitor in (a) and a printed inductor in (b) are utilized. The unit cell area in the $xy$-plane is $\SI{4}{\mm}$~$\times$~$\SI{4}{\mm}$. }
\label{fig2}
\end{figure}

\begin{equation} \label{eq5}
    \phi(x) = \text{sgn}(f)\frac{2\pi}{\lambda}\left(\sqrt{x^2 + f^2}-\lvert f \rvert\right),
\end{equation}
where $f$ is the focal length of a lens and $x$ is the position from the center of the lens.
For sgn$(f)$, $+$ and $-$ are specified for the converging and diverging lens.
The HMS unit cell contains co-located orthogonal electric and magnetic dipole moments to satisfy the boundary conditions for a desired wave transformation.
The electric and magnetic dipoles in the unit cell are represented by a scalar surface electric impedance and magnetic admittance for TE polarized fields~\cite{Vasileios:2021MTTS}.

In this work, we make use of a wire-loop topology to design the HMS unit cells~\cite{Wong:2014Photonics}.
The unit cells have certain surface impedances and admittances for them to synthesize the required $\mathrm{S}_{21}$ phase angles.
The surface impedances of the wire and loop are determined by printed capacitors or inductors in each structure.
Fig.~\ref{fig2} describes the HMS unit cell layout, which consists of physical wire and loop structures with simulation parameters for a Rogers RO3003 substrate (thickness = $\SI{1.524}{\mm}$, $\varepsilon_r=3$, and $\tan \delta = 0.0013$) and 2929 bondply (thickness = $\SI{0.0508}{\mm}$ and $\varepsilon_r = 2.94$).
The thickness of the copper layer on the substrates to form the wire and loop structures is $\SI{18}{\um}$.
The wire structure is shaped by a printed capacitor or inductor on the inner layer of one of the substrates.
The loop structure is formed by printed capacitors on the outer layers of the two substrates with two vias. 
The two substrates are attached by a bonding material.
The area of the unit cell surface is $\SI{4}{\mm}$~$\times$~$\SI{4}{\mm}$.

The magnetic field in the $x$ direction only polarizes the loop.
However, the electric field in the $y$ direction excites not only the wire in the middle layer but also the loop.
Therefore, the design procedure is required to set the desired magnetic admittance of the unit cell by tuning the printed capacitor in the loop, and then the electric impedance is adjusted by tuning the printed capacitor or inductor in the wire~\cite{Wong:2014Photonics}.

\begin{figure}[!t]
\centering
\includegraphics{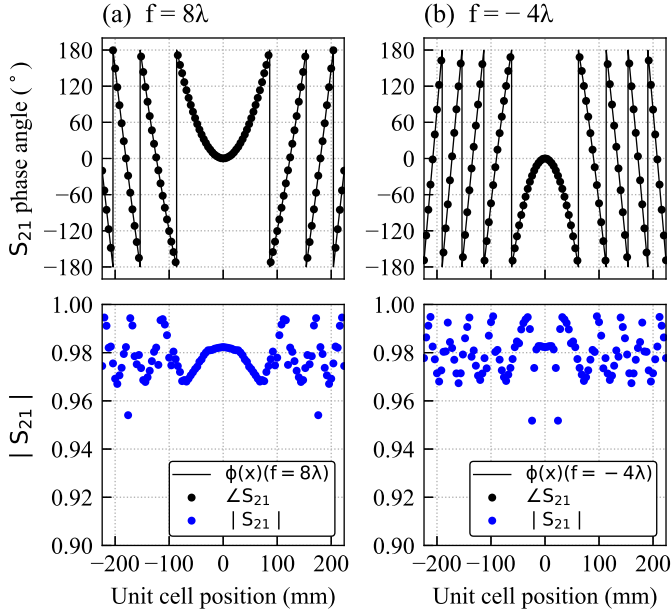}
\caption{Phase angle and magnitude of $\mathrm{S}_{21}$ for each unit cell forming the converging lens ($f=8\lambda$) in (a) and the diverging lens ($f=-4\lambda$) in (b). The simulated phase angle of $\mathrm{S}_{21}$ (black dots) matches the desired phase angle calculated by~\eqref{eq5}. The simulated magnitude of $\mathrm{S}_{21}$ is expressed with blue dots.}
\label{fig3}
\end{figure}

Based on the design procedure above, a wire-loop unit-cell library covering the entire $\mathrm{S}_{21}$ phase range is created.
The unit cells were synthesized by the full-wave electromagnetics solver CST microwave studio.
In calculation, infinite periodic array analysis with these unit cells was utilized.
Fig.~\ref{fig3}(a) and (b) show the magnitude and phase of $\mathrm{S}_{21}$ of the HMS unit cells at the corresponding unit cell position for the converging lens ($f=8\lambda)$ and the diverging lens ($f=-4\lambda$), respectively.
Moreover, the required phase profile, $\phi(x)$, for the two lenses is shown with the black curve.
As shown  $\lvert \mathrm{S}_{21}\rvert$ of the HMS unit cells for the converging lens and the diverging lens are 0.97 and 0.98 on average, respectively.

\begin{figure}[t]
\centering
\includegraphics[width=2.9in]{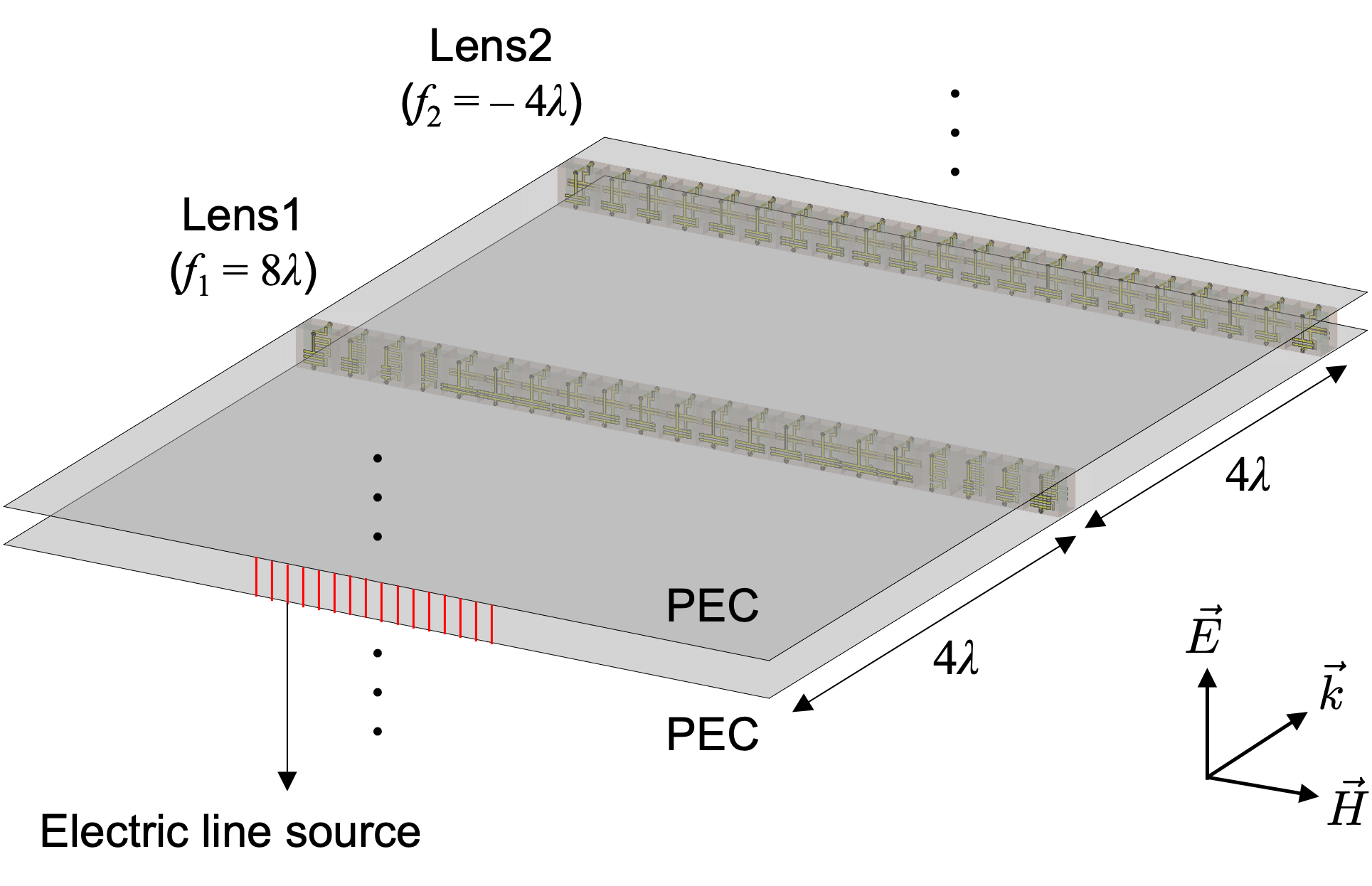}
\caption{The simulation structure with the location of the lenses and the source array, and boundary condition settings.}
\label{fig4}
\end{figure}

\begin{figure}[t]
\centering
\includegraphics{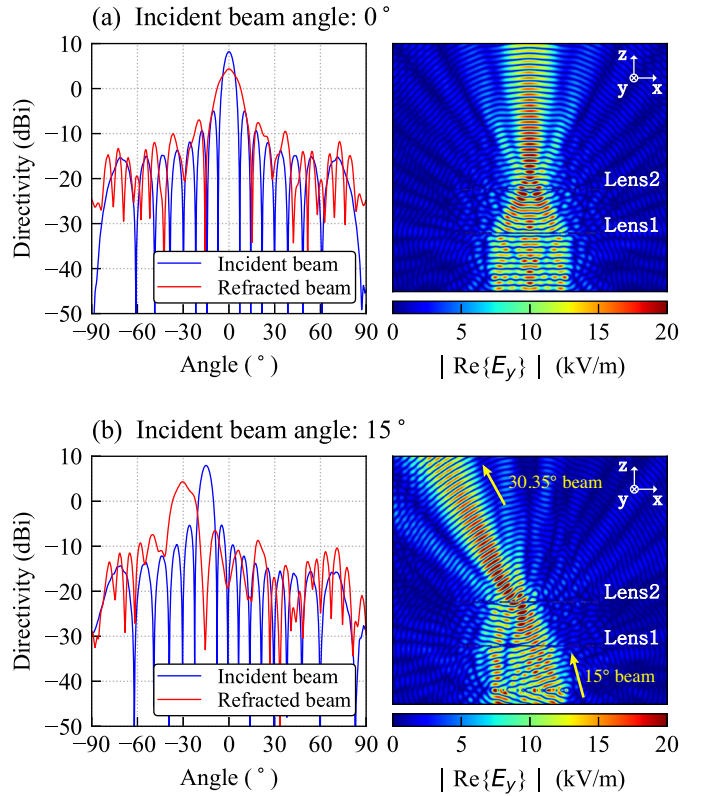}
\caption{Full-wave simulation results: the far-field radiation pattern of the incident beam and the refracted beam (left), and the absolute value of the refracted electric field in the $y$ direction, $\lvert \mathrm{Re}\{ E_y\} \rvert$ (right). (a) Incident beam angle is $\ang{0}$. (b) Incident beam angle is $\ang{15}$.}
\label{fig5}
\end{figure}

\section{Full-wave simulation results of the two-lens HMS system}
In this section, the designed two-lens HMS system as an angle-doubler with the wire-loop unit cells is full-wave simulated by CST microwave studio.
In the simulation, the realized periodicity of the 1D lens array and the source array are in the $y$ direction. Moreover, horizontal upper and lower perfect electric conductor (PEC) walls were set at $y=2$~$\si{mm}$~and~$-2$~$\si{mm}$ as  boundary conditions. 
The specific simulation setup is shown in Fig.~\ref{fig4}.
Fig.~\ref{fig5}(a) depicts the radiation pattern of the incident beam and the refracted beam by the two-lens system in the H-plane and the electric field distribution $\lvert \mathrm{Re}\{ E_y\} \rvert$ when the incident beam is at broadside.
The transmission efficiency of the two-HMS lens system defined as the ratio of the normally transmitted beam power passing through the two lenses to the incident beam power at broadside is $85\%$.

\begin{figure}[t]
\centering
\includegraphics{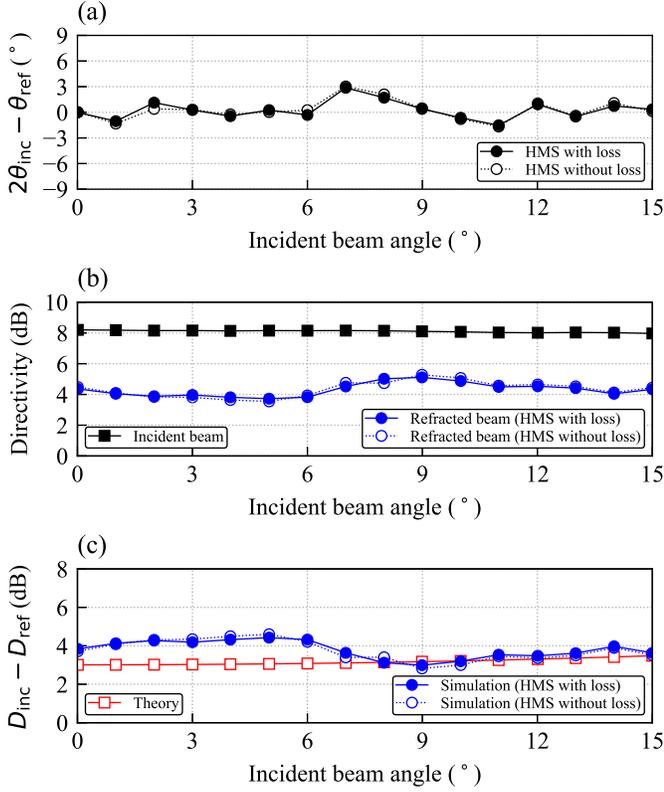}
\caption{(a) The scan error ($\mathrm{2\theta_{inc}} - \mathrm{\theta_{ref}}$) of the two-HMS lens angle doubling system. (b) The directivities of the incident beam (black square solid line) and the refracted beam by the two-HMS lenses at each $\theta_{\mathrm{ref}}$ (blue circle solid line). (c) Directivity degradation ($D_{\mathrm{inc}}-D_{\mathrm{ref}}$) in the simulation (blue circle solid line) calculated with the values in (b) and in the theory (red square solid line) calculated by~\eqref{eq6}. The simulation of the HMSs with losses utilize the material constants of copper and  Rogers RO3003. The simulation of the HMSs without losses, whose results are shown with dotted lines in the three plots, is conducted with PEC and lossless dielectric materials.}
\label{fig6}
\end{figure}

In addition, the reflectance of the broadside beam to the system is $3.4\%$.
As shown, at broadside the incident beam from the phased array becomes a diverging beam with a directivity degradation of $\SI{3.8}{\dB}$.
Note that theoretically the directivity reduction when passing through the two-lens system is given by \eqref{eq6}~\cite{Gleb:two_lens}.
\begin{equation} \label{eq6}
    \frac{D_{\mathrm{ref}}(\theta_{\mathrm{ref}})}{D_{\mathrm{inc}}(\theta_{\mathrm{inc}})} = \frac{1}{\alpha} \cdot \frac{\cos{\theta_{\mathrm{ref}}}}{\cos{\theta_{\mathrm{inc}}}},
\end{equation}
where $D_\mathrm{ref}$ is the directivity of the beam passing through the two lenses, $D_\mathrm{inc}$ is the directivity of the beam from the source array, and $\alpha$ is the angular scan enhancement factor. 
Accordingly, for angle doubling, a from $3$ to $\SI{3.5}{\dB}$ degradation of the directivity of this system is unavoidable, when the incident beam angle is between $\ang{-15}$ and $\ang{15}$.
Fig.~\ref{fig5}(b) shows the same plot as Fig.~\ref{fig5}(a), but the incident beam angle is changed to $\ang{-15}$ off broadside.
It is shown that the beam angle is extended from $\ang{-15}$ to $\ang{-30.35}$ by the two HMS lenses.
Furthermore, the directivity difference between the incident beam and the refracted beam is $\SI{3.63}{\dB}$, which is in good agreement with the expected degradation at this incident angle, which is $\SI{3.48}{\dB}$ by \eqref{eq6}.

\begin{figure}[t]
\centering
\includegraphics{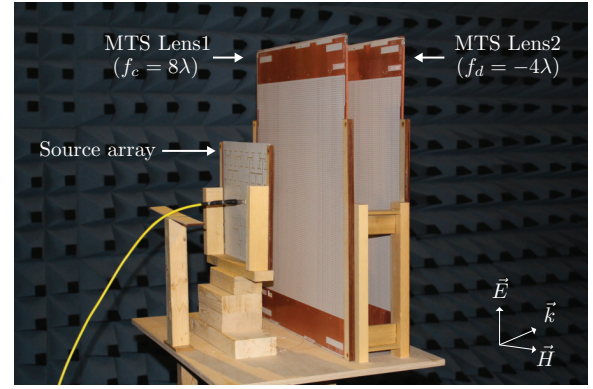}
\caption{Experimental setup, showing the source array illuminating the two lenses. The E-plane lies vertically as shown in the vector diagram.}
\label{fig7}
\end{figure}

\begin{figure}[t]
\centering
\includegraphics{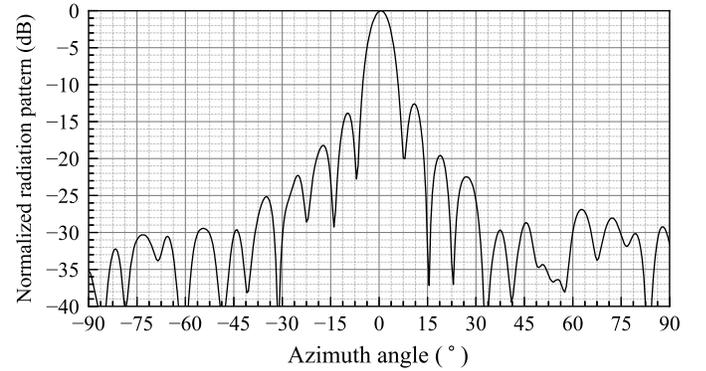}
\caption{Measured source array pattern with the lenses removed (see Fig.~\ref{fig7}).}
\label{fig8}
\end{figure}

Lastly, the angle doubling performance of the two-HMS lens system at various incident beam angles is analyzed.
Fig.~\ref{fig6}(a) depicts that the two-HMS lens system performs well as an angle doubler by showing that the magnitude of the scan error defined as $\lvert\mathrm{2\theta_{inc}} - \mathrm{\theta_{ref}}\rvert$ is less than $\ang{2.9}$ when the incident angles are below $\ang{\pm15}$.
Fig.~\ref{fig6}(b) describes the peak directivities of the incident beam from the phased array and the refracted beam passing through the angle-doubler.
Fig.~\ref{fig6}(c) shows that the directivity degradation at incident beam angles below $\ang{15}$ is $3.7\pm0.7$~$\si{dB}$ by the simulation.
It should be noted that there is a non-negligible difference between simulation and theory at smaller incident angles below $\ang{6}$ in Fig.~\ref{fig6}(c).
This discrepancy might be due to the fact that the physical unit cells of the two lenses have a non-uniform transmittance spatially, in which the difference is $8\%$ at most, as inferred in Fig.~\ref{fig3}.
In fact, the simulation results with the analogous system using two dielectric lenses show similar directivity degradation to the theory below $\ang{10}$ of the incident angle.
It is also worth noting that material losses of the HMSs do not affect the directivity degradation, as described in Fig.~\ref{fig6}. Namely, the simulation results conducted with ideal materials such as PEC and lossless dielectric substrate ($\tan \delta = 0$) show very little difference from those with copper and a lossy dielectric substrate (Rogers RO3003).
Here, the simulated power loss through the two lenses is $11.6\%$.

\begin{figure}[t]
\centering
\includegraphics{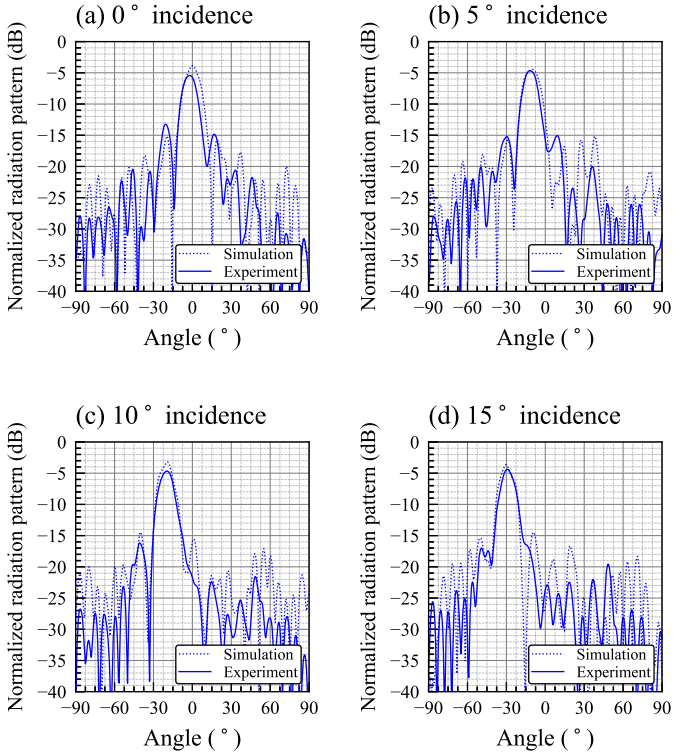}
\caption{Comparison of measured and simulated angle-doubled patterns at $\SI{10}{\GHz}$. Angle doubling is observed, and experiment closely resembles the simulations.}
\label{fig:SimExpComparison}
\end{figure}

\begin{figure}[t]
\centering
\includegraphics{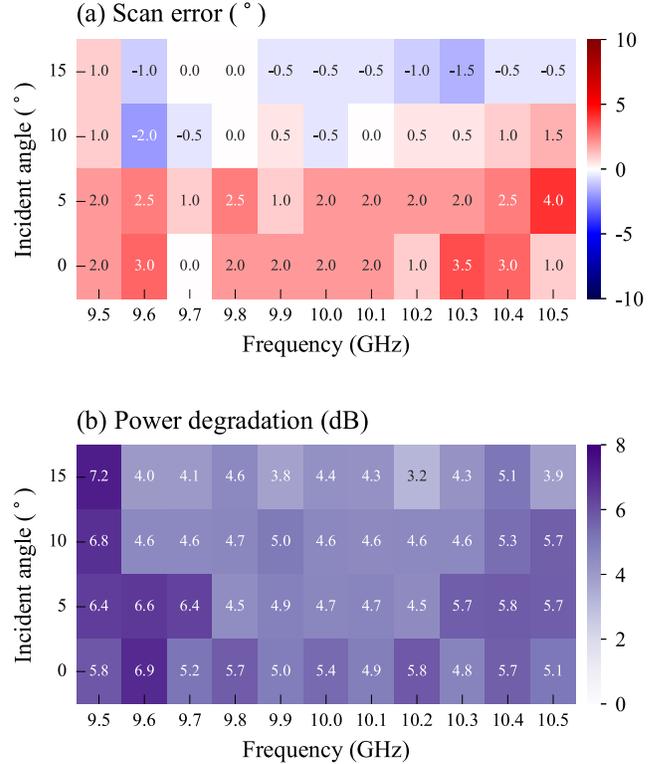}
\caption{Complete measured scan error and power degradation data for the two-lens doubler.}
\label{fig:FullData}
\end{figure}

\section{Experimental results}
Two PCB metasurface lenses of size $\SI{452}{\mm}$~$\times$~$\SI{452}{\mm}$ were fabricated following the validation by full-wave simulation.
The designed metasurface lenses were repeated $113$~times, which is $15\lambda$ in length at $\SI{10}{\GHz}$, along E-plane.
To demonstrate the enhancement of the scan angle range of a phased array with the two metasurface lens system, an anechoic chamber antenna test was performed.
Fig.~\ref{fig6} shows the overall two-lens system with a source array for incident beams in our anechoic chamber.
For the incident beam source, we have designed and fabricated a $16\times16$ patch antenna array with element spacing of $\lambda/2$, in which the patches were aperture coupled to the corporate feed network.
This array produced a beam at broadside, and the patch array was rotated around its center in order to experimentally produce off-broadside beams.
Fig.~7 shows measured patterns of the array at $\SI{10}{\GHz}$.

Fig.~\ref{fig:SimExpComparison} compares measured and simulated patterns at the particular frequency of $\SI{10}{\GHz}$.
To compare the experimental and simulation data (for which some subtle differences exist) and because we are interested in the directivity degradation, the plotted patterns are normalized to the peak of the source of the particular scenario being plotted -- for example, in part (b) the $\ang{5}$ incidence scenario is plotted, for which the array is tilted to $\ang{5}$ but still has the same peak while in simulation the array is steered and thus has a different peak.
Clearly, the experimental setup indeed achieves angle-doubling with little error. For completeness, Fig.~\ref{fig:FullData} shows obtained (a) scan error and (b) power degradation for all measured values of $f$ and $\theta_\mathrm{inc}$. 
It appears that the angle doubler performs the best for frequencies within $10\pm0.2$~$\si{\GHz}$. 
The scan error is small, and the power degradation is $\SI{4.7}{\dB}$ on average. 
These values suggest that the losses, non-ideal reflections and scattering all amount to around $\SI{1.7}{\dB}$ degradation in directivity (received power, if speaking in terms of the experiment), which is expected.


\section{Conclusion}
A physical design for a two-Huygens' metasurface lens system to double the scan angle of a phased array has been numerically and experimentally demonstrated.
Design parameters for the two-HMS angle doubler are obtained based on ray optics analysis.
The HMS lenses are designed for a quadratic-like $\mathrm{S}_{21}$ phase profile, and the wire-loop topology is deployed to implement their unit cells.
The HMS-lens doubler is placed in the near-field of the phased array leading to a compact architecture.
The performance of the system has been verified by full-wave simulations and experiments.
First of all, the magnitude of $\mathrm{S}_{21}$ of the unit cells we utilize for the converging lens and the diverging lens are 0.97 and 0.98 on the average, respectively, covering the entire $\mathrm{S}_{21}$ phase angles.
Furthermore, the proposed two-lens HMS angle doubler functions properly showing that the scan angle of a uniform phased array is enhanced by a factor of two when the angle of incidence is between $\ang{-15}$ and $\ang{15}$ which was our design specification.
The simulation results show that the directivities of the refracted beams by the two-HMS lens system are degraded by nearly $\SI{3.7}{\dB}$, in good agreement with theory.
Experimental results also match well with the simulation by showing that angle-doubling is well achieved with low scan error, $\ang{\pm2}$, and around $\SI{4.7}{\dB}$ power degradation at $\SI{10}{\GHz}$.
Finally, it can be pointed out that if the proposed two-lens HMS system is reversed, it can function as a beam expander to enhance the directivity of a feeding antenna~\cite{Dorrah:2019URSI}.


%



\section*{Acknowledgment}
Computations were performed on the Niagara supercomputer at the SciNet HPC Consortium. SciNet is funded by: the Canada Foundation for Innovation; the Government of Ontario; Ontario Research Fund - Research Excellence; and the University of Toronto.

\ifCLASSOPTIONcaptionsoff
  \newpage
\fi

\end{document}